\documentclass[amsmath,amssymb,superscriptaddress,nobalancelastpage,prx,twocolumn]{revtex4-1}
\usepackage{blindtext}
\usepackage{graphicx}
\usepackage{color}
\usepackage{changes}

\newcommand{\Bacu}{Ba$_2$Cu$_3$O$_4$Cl$_2$}
\newcommand{\SCOC}{Sr$_2$CuO$_2$Cl$_2$}

\begin{document}
\date{\today}
\title{Electronic and magnetic excitations in the ``half-stuffed" Cu--O 
planes of \Bacu~measured by resonant inelastic x-ray scattering}

\author{S. Fatale}
\affiliation{Institute of Physics (IPHYS), \'{E}cole Polytechnique 
F\'{e}d\'{e}rale de Lausanne (EPFL), CH-1015 Lausanne, Switzerland}

\author{C. G. Fatuzzo}
\affiliation{Institute of Physics (IPHYS), \'{E}cole Polytechnique 
F\'{e}d\'{e}rale de Lausanne (EPFL), CH-1015 Lausanne, Switzerland}

\author{P. Babkevich}
\affiliation{Institute of Physics (IPHYS), \'{E}cole Polytechnique 
F\'{e}d\'{e}rale de Lausanne (EPFL), CH-1015 Lausanne, Switzerland}

\author{N. E. Shaik}
\affiliation{Institute of Physics (IPHYS), \'{E}cole Polytechnique 
F\'{e}d\'{e}rale de Lausanne (EPFL), CH-1015 Lausanne, Switzerland}

\author{J. Pelliciari$^*$}
\affiliation{Swiss Light Source, Paul Scherrer Institut, CH-5232 Villigen 
PSI, Switzerland}
\email{Present address: Department of Physics, Massachusetts Institute of 
Technology, Cambridge, MA 02139, USA}

\author{X. Lu}
\affiliation{Swiss Light Source, Paul Scherrer Institut, CH-5232 Villigen 
PSI, Switzerland}

\author{D. E. McNally}
\affiliation{Swiss Light Source, Paul Scherrer Institut, CH-5232 Villigen 
PSI, Switzerland}

\author{T. Schmitt}
\affiliation{Swiss Light Source, Paul Scherrer Institut, CH-5232 Villigen 
PSI, Switzerland}

\author{A. Kikkawa}
\affiliation{RIKEN Center for Emergent Matter Science (CEMS), Wako, Saitama 
351-0198, Japan}

\author{Y. Taguchi}
\affiliation{RIKEN Center for Emergent Matter Science (CEMS), Wako, Saitama 
351-0198, Japan}

\author{Y. Tokura}
\affiliation{RIKEN Center for Emergent Matter Science (CEMS), Wako, Saitama 
351-0198, Japan}
\affiliation{Department of Applied Physics, University of Tokyo, Bunkyo-ku 
113-8656, Japan}

\author{B. Normand}
\affiliation{Laboratory for Neutron Scattering and Imaging, Paul Scherrer 
Institut, CH-5232 Villigen PSI, Switzerland}

\author{H.~M.~R\o{}nnow}
\affiliation{Institute of Physics (IPHYS), \'{E}cole Polytechnique 
F\'{e}d\'{e}rale de Lausanne (EPFL), CH-1015 Lausanne, Switzerland}

\author{M. Grioni}
\affiliation{Institute of Physics (IPHYS), \'{E}cole Polytechnique 
F\'{e}d\'{e}rale de Lausanne (EPFL), CH-1015 Lausanne, Switzerland}
	
\begin{abstract}
We use resonant inelastic x-ray scattering (RIXS) at the Cu L$_3$ edge to 
measure the charge and spin excitations in the ``half-stuffed'' Cu--O planes 
of the cuprate antiferromagnet Ba$_2$Cu$_3$O$_4$Cl$_2$. The RIXS line shape 
reveals distinct contributions to the $dd$ excitations from the two 
structurally inequivalent Cu sites, which have different out-of-plane 
coordinations. The low-energy response exhibits magnetic excitations. We 
find a spin-wave branch whose dispersion follows the symmetry of a CuO$_2$ 
sublattice, similar to the case of the ``fully-stuffed'' planes of tetragonal 
CuO (T-CuO). Its bandwidth is closer to that of a typical cuprate material, 
such as \SCOC, than it is to that of T-CuO. We interpret this result as 
arising from the absence of the effective four-spin inter-sublattice 
interactions that act to reduce the bandwidth in T-CuO.
\end{abstract}

\maketitle
\section{Introduction}

The electronic properties of cuprate materials are determined largely by 
their strongly correlated Cu--O planes, which are assembled from CuO$_4$ 
units (plaquettes), typically connected in a corner-sharing arrangement, 
as shown in Fig.\,1(a). The undoped parent compounds are charge-transfer 
insulators \cite{ZSA}. In the ionic limit the Cu ions have a $d^9$ 
configuration, with a single $3d$ hole in the $3d_{x^2-y^2}$ orbital, while 
the O $2p$ band is completely filled. The gap is defined by excitations 
from the O $2p$ to the Cu $3d$ band: $d^9 \rightarrow d^{10} \underline{L}$, 
where $\underline{L}$ represents a hole in the ligand (oxygen) band. The 
magnetic coupling between the local $S = 1/2$ moments on the plaquettes is 
antiferromagnetic (AFM), with a typical energy scale $J$ of order 0.1~eV.
When holes are introduced by chemical doping, or by removing one electron 
in an angle-resolved photoemission (ARPES) experiment, the lowest-energy 
state is not the triplet expected from the Hund rules, but rather a 
``Zhang-Rice singlet'' (ZRS), a symmetry-adapted $d^9 \underline{L}$ 
superposition of a $3d_{x^2-y^2}$-hole and an O hole delocalized on the 
4 O ions of the plaquette \cite{ZRS,EskesSaw}.

Much experimental and theoretical work has been devoted to the study of the 
properties of the ZRS, with ARPES playing a major role \cite{Damascelli}. The 
magnetic excitations have also been studied extensively, both in theory and in 
experiment, with the latter investigations performed primarily by inelastic 
neutron scattering (INS) \cite{HaydenReview}. A recent focus in the cuprates 
community has been to bridge the differences between some of the very different 
experimental techniques, in order to obtain a consistent description of the 
band structure, optical, and magnetic response of a single material using a 
single set of electronic parameters. These parameters would vary systematically 
across the different families of cuprates according to factors such as 
coordination, layering, and the role of apical O atoms. In this context, 
resonant inelastic x-ray scattering (RIXS) has emerged as a new probe of 
both charge and magnetic excitations in the cuprates, offering the possibility 
of mapping both crystal-field splittings and the full magnon spectrum, even 
with only rather small single-crystal samples \cite{SaraReview}.

The corner-sharing arrangement of plaquettes is quite ubiquitous in cuprates 
containing Cu--O planes. By contrast, the alternative of edge-sharing 
coordination is, to a large extent, known only in quasi-1D compounds, such as 
CuGeO$_3$ or Li$_2$CuO$_2$ \cite{Monney}. A notable exception is tetragonal
CuO (T-CuO), which can be grown as a thin film on an STO substrate 
\cite{Siemons}. This  metastable form of CuO contains square-lattice 
Cu$_2$O$_2$ planes [Fig.\,1(b)], built from edge-sharing CuO$_4$ plaquettes, 
which are stacked along the $c-$axis. Each plane can be considered as a 
superposition of two identical corner-sharing CuO$_2$ sublattices (which 
we denote as Cu$_A$ and Cu$_B$) with common O ions, displaced by $(0.5\,a, 
0.5\,a)$ with respect to each other. It can also be seen as a CuO$_2$ 
lattice where an additional Cu ion has been added at the center of each 
unit cell, also known as the ``fully-stuffed'' configuration.

The electronic structure of T-CuO has been measured by ARPES \cite{Moser2014}. 
It exhibits ZRS-type quasiparticles that are compatible with a moderately weak 
coupling between the two sublattices. RIXS measurements have shown that the 
spin-wave dispersion of T-CuO follows the symmetry of the magnetic Brillouin 
zone (BZ) of a single AFM CuO$_2$ sublattice \cite{Moser2015}, which is 
represented in Fig.\,1(d). The momentum-dependence is found to be qualitatively 
the same as for typical insulating cuprates such as \SCOC~(SCOC) \cite{Guarise},
with features including a weak dispersion along the $(\pi/2,\pi/2)\rightarrow 
(\pi,0)$ boundary of the magnetic BZ, which has been interpreted within a 
pure $S = 1/2$ model as being indicative of longer-range interactions beyond 
the simple nearest-neighbor (NN) Heisenberg model \cite{Coldea,Guarise}. 
However, the magnon energies in T-CuO are approximately 30\% smaller than in 
SCOC. This suggests one of two obvious possibilities: either there are 
competing magnetic interactions between the Cu spins of the fully-stuffed 
lattice or, given that the exchange energy, $J$, scales with the electron 
hopping amplitude, $t$, the different crystal structure of T-CuO results in 
a different hopping \cite{Moser2015}. Because the two sublattices of T-CuO 
are identical, the quantitative analysis is not free from ambiguities. In 
particular, the degeneracy of the magnetic modes is lifted by the weak 
inter-sublattice coupling, but the resulting splitting is too small to be 
resolved with the present energy resolution of RIXS.

\begin{figure}[t]
\begin{center}
\includegraphics[scale=0.7]{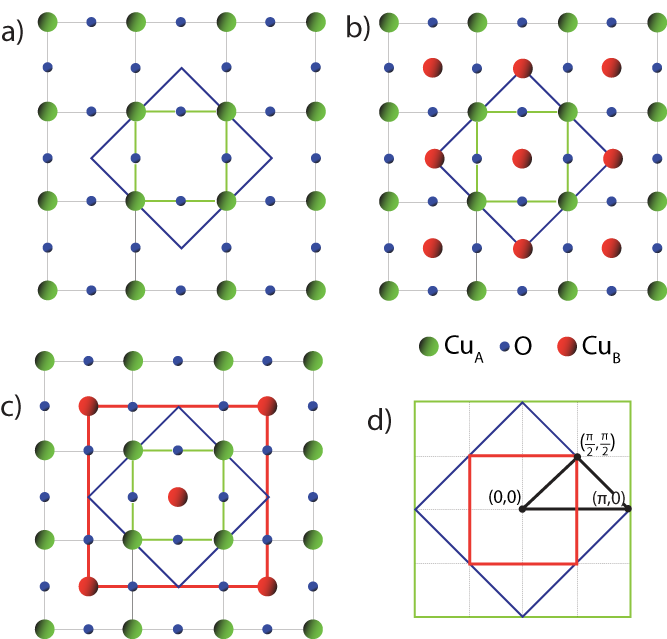}
\caption{(a) The CuO$_{2}$ plane of a corner-sharing cuprate. 
(b) The ``fully-stuffed" Cu$_2$O$_2$ plane of T-CuO, with identical A (green) 
and B (red) sublattices. In both (a) and (b), the green and blue squares are 
respectively the structural and magnetic unit cells. (c) The ``half-stuffed'' 
Cu$_3$O$_4$ plane of Ba-2342 with inequivalent Cu$_A$ (green) and Cu$_B$ (red) 
sublattices. The blue square is the structural unit cell and also the magnetic 
unit cell of the Cu$_A$ sublattice. The red square is the magnetic unit cell 
below T$_{N,B}$, where both sublattices are ordered. (d) BZs corresponding to 
the same-color unit cells in (a)-(c).}
\label{fig1}
\end{center}
\end{figure}

\Bacu~(Ba-2342) provides an opportunity to investigate an intermediate 
situation between the conventional ``empty'' corner-sharing CuO$_2$ square 
lattice and the fully-stuffed edge-sharing square lattice of T-CuO. 
Ba-2342 contains Cu--O planes with a Cu$_3$O$_4$ stoichiometry, which can 
be seen as the superposition of two corner-sharing CuO$_2$ sublattices 
[Fig.\,1(c)]. However, in contrast to T-CuO, in Ba-2342 the two sublattices 
have different sizes. The lattice parameter ($a = 3.90$\,\AA) of the Cu$_A$ 
sublattice (shown in green in Fig.\,1) is almost identical to that of typical 
cuprates such as SCOC ($3.97$\,\AA). The Cu$_B$ sublattice (shown in red) is 
twice as large. The Cu$_3$O$_4$ plane can be obtained from the fully-stuffed 
Cu$_2$O$_2$ plane of T-CuO by removing every second atom from one of the two 
identical sublattices, and it is therefore often referred to as 
``half-stuffed.'' An ARPES study of Ba-2342 revealed two distinct ZRS bands 
dispersing independently and with different bandwidths on the two sublattices 
\cite{Golden}.

\begin{figure}[t]
\includegraphics[scale=0.4]{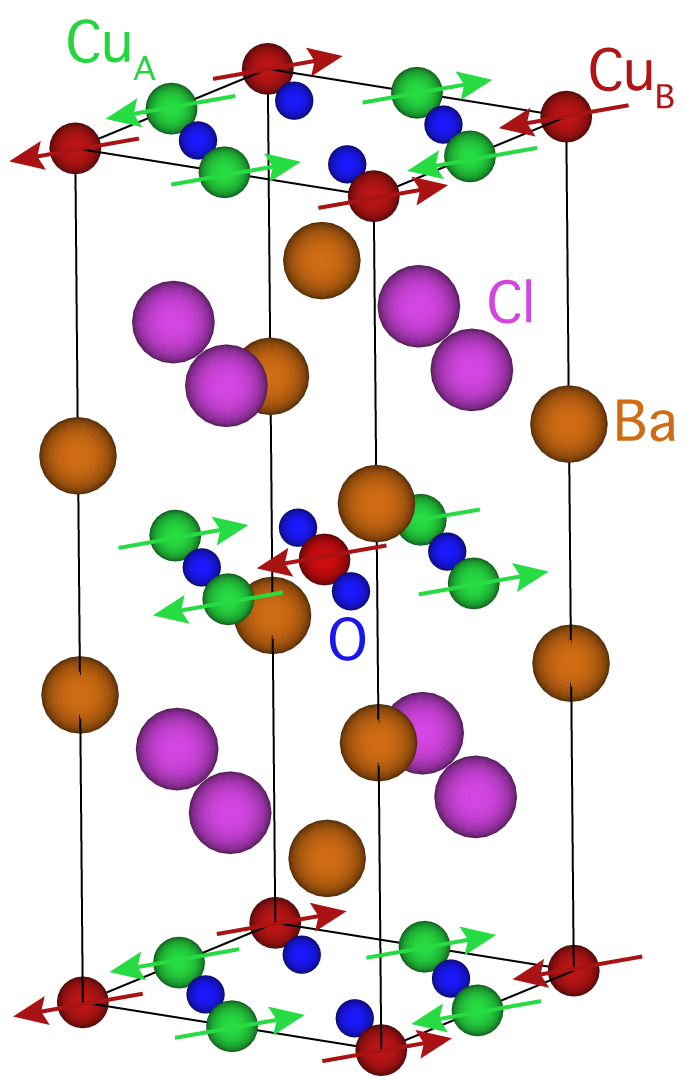}
\caption{The 3D structure of Ba-2342 showing the tetragonal unit cell, whose 
mid-plane coincides with the blue square in Fig.\,1(c). The arrows illustrate 
the magnetic order in the low-temperature phase below T$_{N,B}$.}
\label{fig2}
\end{figure}

The BZs for the different Cu--O planes discussed above are shown in Fig.\,1(d). 
The largest square (green) is the BZ corresponding to the primitive unit cell 
of the SCOC and T-CuO lattices. The smaller and $45^{\circ}$-rotated square 
(blue) is the 2D AFM BZ of SCOC and of T-CuO. It is also the 2D BZ of Ba-2342, 
as well as the AFM BZ of the Cu$_A$ sublattice. The smallest square (red) is 
the AFM BZ of the second (red) sublattice, and also the overall magnetic BZ 
of Ba-2342 in the low-temperature ordered phase.

The Cu$_A$ and Cu$_B$ sublattices order antiferromagnetically at quite 
different temperatures, respectively $T_{N,A} = 324$\,K and $T_{N,B} = 31$\,K 
\cite{Babkevich}, which already implies a weak coupling and quite different 
physics on the two sublattice. Below $T_{N,B}$ the ordered moments, $\sim0.6\,
\mu_B$ on both the Cu$_A$ and Cu$_B$ sites, are collinear and parallel with 
the (1,0) direction, as depicted in Fig.\,2. The spin waves associated with 
the Cu$_A$ and Cu$_B$ sublattices have correspondingly different energy scales, 
of order $300$\,meV for the former and $20$\,meV for the latter. To date, only 
the low-energy part of the dispersion has been studied by INS in the 
isostructural sister compound Sr-2342\cite{Kim}. The full spectrum of Ba-2342 
was measured in a very recent INS experiment \cite{Babkevich}. However, as a 
result of the weak scattering intensity, an accurate determination of the 
magnon dispersion above approximately $250$\,meV by INS is challenging. 

Here we present RIXS measurements of the magnetic excitation spectrum of 
Ba-2342 and compare them with analogous data both from a typical cuprate 
parent compound, for which we use SCOC, and from T-CuO. When applied for 
this purpose, RIXS offers a number of advantages or complementary features 
when compared to INS. While its ability to collect high-quality data on 
very small (sub-mm-sized) samples is not critical in the common cuprates, 
the large RIXS cross-section has a wave-vector dependence quite different 
from INS. It can measure single-magnon bands and multi-magnon continua at 
the same time, and access both Raman and non-Raman features depending on 
the excitation energy. In particular for cuprates, it has far better 
statistics than INS at the upper band-edge energy around 300 meV. It is 
also readily capable of mapping electronic excitations, giving direct 
information about the local crystal environment (crystal-field excitations), 
which are completely inaccessible by INS. 

The structure of this article is as follows. In Sec.~II we discuss the 
details of our experimental analysis. In Sec.~III we present our results, 
which we separate into an analysis (Sec.~IIIA) of the higher-energy $dd$ 
excitations and extraction of the tetragonal crystal-field parameters, 
our measurements (Sec.~IIIB) of the low-energy magnetic excitations, and 
the extraction of the spin-wave spectrum (Sec.~IIIC) combined with some 
theoretical considerations related to the effective modelling of the 
different cuprate planes in SCOC, Ba-2342, and T-CuO. In Sec.~IV we 
present a short summary and conclusion. 

\section{Experiment}

Single crystals of Ba-2342 were grown by the floating zone method, as described 
in Ref.\,\cite{Babkevich}, and were characterized by x-ray powder 
diffraction. For our RIXS measurements, a crystalline plaquette (of approximate 
size $2 \times 2 \times 0.5$\,mm$^3$) was aligned by von Laue diffraction and 
then mounted on a cryogenic manipulator with two angular degrees of freedom. 
This assembly was inserted in the UHV system and cleaved \textit{in situ} by 
the top-post method to expose the $(001)$ surface. The sample was kept at a 
temperature $T = 20$\,K during all measurements. 

Cu L$_3$-edge ($2p_{3/2} \rightarrow 3d$; $h\nu \approx 930$\,eV) RIXS data 
were acquired at the ADvanced RESonant Spectroscopy (ADRESS) beam line at 
the Swiss Light Source (SLS), which is located at the Paul Scherrer Institut 
\cite{beamline}. The scattering angle was set to 130$^{\circ}$ in the horizontal 
plane, which contained the $c$ axis of the sample [Fig.\,3(a)]. The incident 
light was $\pi$-polarized. The energy resolution estimated from the elastic 
peak of a coplanar polycrystalline Cu sample was $130$ meV, which sets the 
measured peak widths. However, the accuracy to which the peak centers may be 
located is of order 40 meV over much of the BZ. The total transferred momentum 
was $Q = 0.85$\,\AA$^{-1}$. By rotating the sample around the vertical axis, 
its projection $q$ on the $ab$ plane could be varied between $-0.76$\,\AA$^{-1}$ 
\,(grazing incidence) and $0.76$\,\AA$^{-1}$ \,(grazing emission), with $q = 0$ 
corresponding to specular geometry. Following convention, the spectra were 
normalized to the same integrated intensity in the $dd$ manifold.
Self-absorption, i.e.~the absorption of the scattered beam by the sample, is 
significant only in the quasielastic region and near grazing emission, where 
it affects the intensities of the spectral features. Self-absorption-induced 
energy shifts are too small to be resolved with the present energy resolution 
and are not considered in the discussion to follow.

\begin{figure}[t]
\includegraphics[scale=0.65]{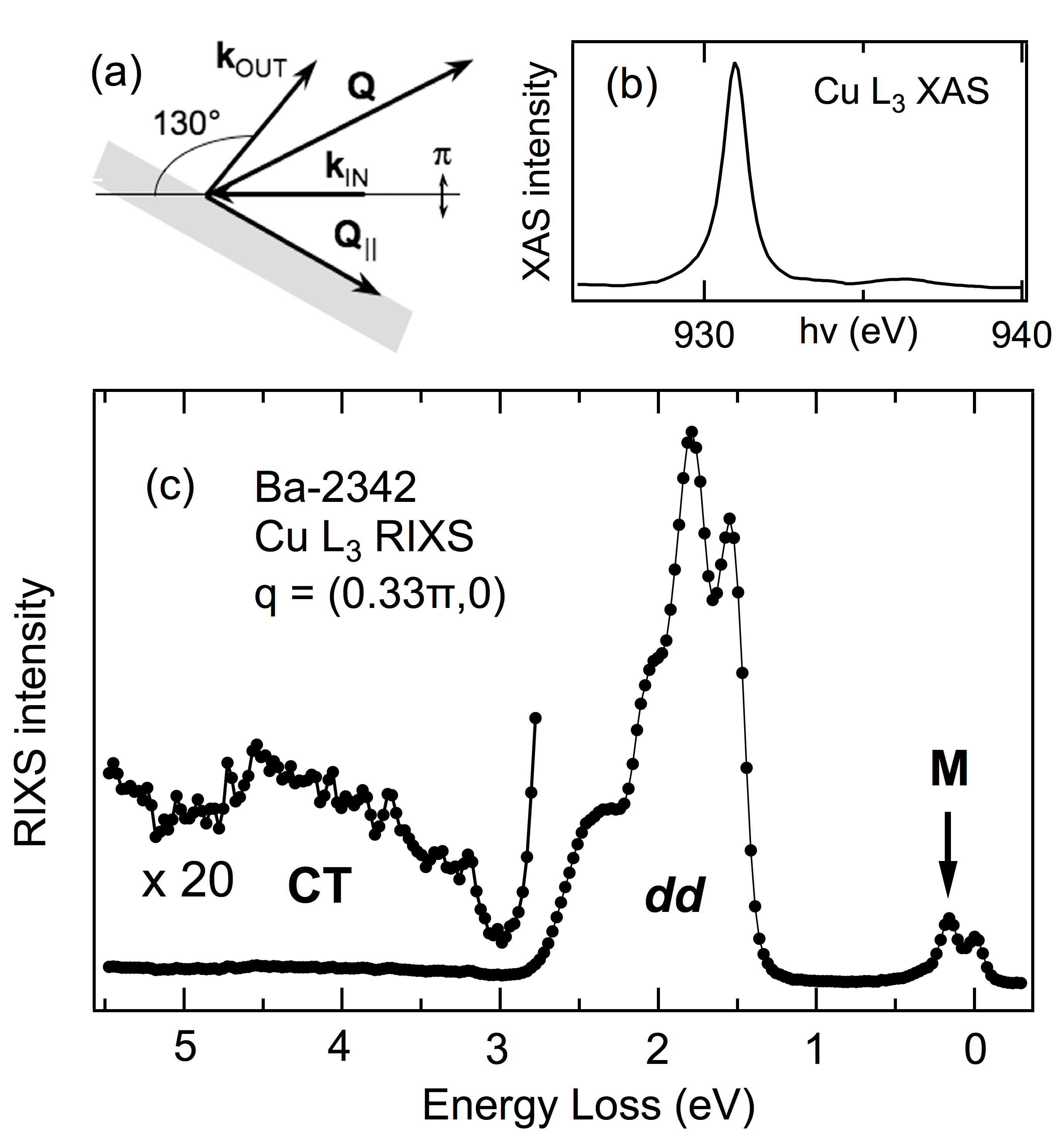}
\caption{(a) Experimental geometry. The scattering plane is horizontal and 
the sample was rotated around a vertical axis. (b) Cu L$_3$ XAS spectrum. (c) 
A typical RIXS spectrum for Ba-2342, measured as a function of the energy loss, 
($h\nu_{in} - h\nu_{out}$). The elastic peak is at $E = 0$. The magnon feature 
(M), the $dd$, and the charge-transfer (CT) manifolds are indicated. A 
magnified ($\times 20$) version of the weak CT signal is included.}
\label{fig3}
\end{figure}
\begin{table*}[!t]
\begin{tabular}{l|c|c|c|c|c|c}
 & $d_{Cu-O}$ (\AA) & $d_{Cu-X}$ (\AA) & E$_{3z^2-r^2}$ (eV) & E$_{xy}$ (eV) & 
E$_{xz/yz}$ (eV) & Ref.\\
\hline
\hline
La$_2$CuO$_4$ & 1.90 & 2.43 & 1.70 & 1.80 & 2.12 & \cite{Moretti} \\
T-CuO & 1.95 & 2.67 & 1.75 & 1.6 & 1.5 & \cite{Moser2015} \\
Sr$_2$CuO$_2$Cl$_2$ & 1.98 & 2.86 & 1.97 & 1.5 & 1.84 & \cite{Moretti} \\
Ba$_2$Cu$_3$O$_4$Cl$_2$ & 1.95 & 3.43/4.94 & 2.04/2.44 & 1.57 & 1.78 & this 
work \\
CaCuO$_2$ & 1.93 & $\infty$ & 2.65 & 1.65 & 1.95 & \cite{Minola} \\
 \hline
\end{tabular}
\caption{In-plane Cu-O ($d_{Cu-O}$) and out-of-plane Cu-anion ($d_{Cu-X}$) 
separations, shown together with the energies of a hole in the various Cu
$3d$ orbitals relative to the $3d_{x^2-y^2}$ orbital, as measured by RIXS in 
representative cuprate materials. For Ba-2342, values for both the Cu$_A$ 
and Cu$_B$ sites are indicated.}
\end{table*}

\section{Results and Analysis}
RIXS is a second-order coherent optical process, initiated by the resonant 
absorption of a photon by a core-level electron. Specifically, Cu L$_3$ RIXS 
from a  divalent (Cu$^{\textrm{II}}$) copper ion is described schematically as 
$2p_{3/2}^4\,d^9 + h\nu_{in} \rightarrow 2p_{3/2}^3\,d^{10} \rightarrow 2p_{3/2}^4 \,
(d^9)^* + h\nu_{out}$, where $(d^9)^*$ denotes the ground state or an excited 
state of the $3d^9$ configuration while $h\nu_{in}$ and $h\nu_{out}$ are the 
energies of the incident and scattered photons. Figure 3(c) displays a 
representative RIXS spectrum measured on Ba-2342, which is shown as a 
function of the photon energy loss, $(h\nu_{in} - h\nu_{out})$. Here and in 
the following, the incident energy is tuned to the peak of the L$_3$ x-ray 
absorption spectrum, which is shown in Fig.~3(b).
The prominent feature of the RIXS line shape around $2$\,eV is the $dd$ 
manifold, representing optical transitions from the occupied crystal-field 
levels to the empty $3d_{x^2-y^2}$ orbital. These $\Delta L = 0$ transitions are 
dipole-forbidden but become allowed in the second-order RIXS process. The peak 
at $E = 0$ contains the elastic peak, corresponding to transitions that bring 
the system back to the ground state, and has a quasielastic tail arising from 
unresolved phonon excitations. Peak M is the signature of a collective spin 
excitation, which is enabled by the strong (of order 20\,eV) spin-orbit 
interaction in the intermediate $2p$ core-hole state. The weak, broad CT 
feature above the $dd$ manifold is a continuum of charge-transfer final 
states, mostly of $d^{10} \underline{L}$ character. These states couple to 
the $2p_{3/2}^3\,d^{10}$ intermediate state through their minority $d^9$ 
component and their spectral weight is therefore quite small in Cu L$_3$ 
RIXS, whereas they dominate the RIXS spectral function at the oxygen K 
($1s$) edge \cite{Harada}.

\begin{figure}[b]
\includegraphics[scale=0.4]{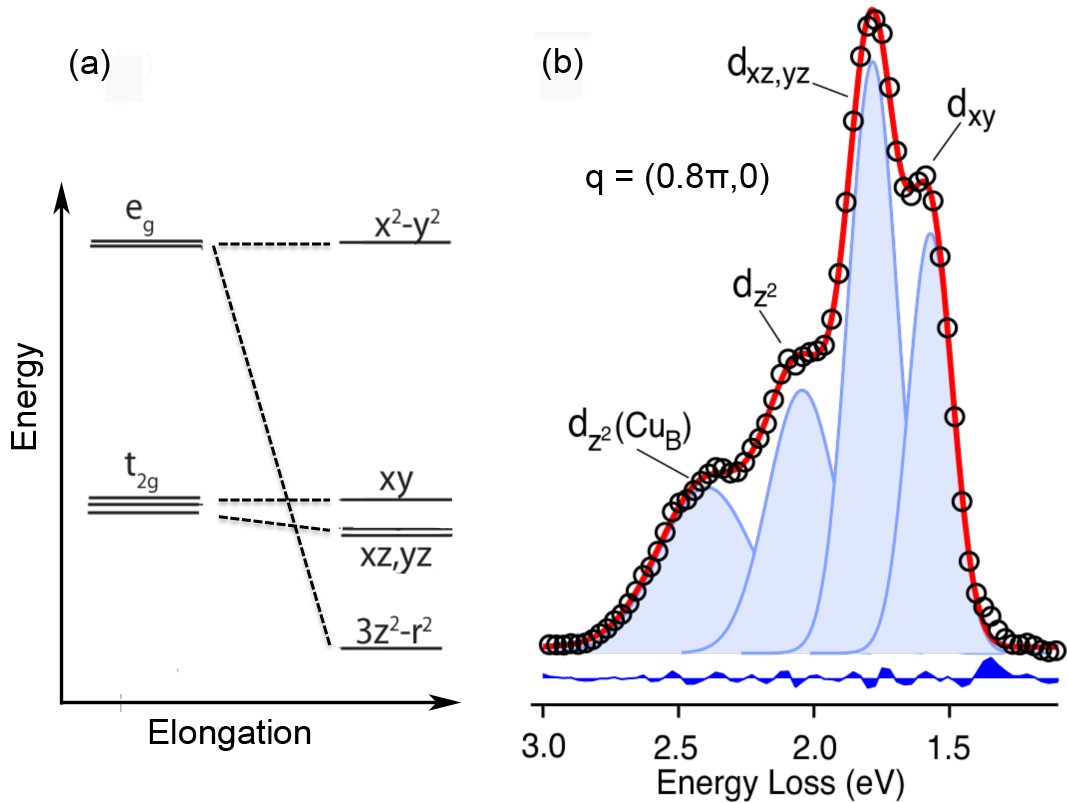}
\caption{(a) Effect of a tetragonal distortion (elongation) on the energies 
of the Cu $3d$ states. Energies are relative to the $3d_{x^2-y^2}$ orbital. 
(b) Fit of the $dd$ line shape of a representative spectrum to four Gaussian 
peaks. The blue line at the bottom is the residue.} 
\label{fig4}
\end{figure}

\subsection{$dd$ excitations}

We first consider the $dd$ excitations, which contain information concerning 
the local environment of the Cu ions. For a Cu$^{\textrm{II}}; d^9$ ion in an 
undistorted octahedral site (cubic symmetry), a single peak would be observed 
at an energy equal to the separation between the $e_g$ and $t_{2g}$ states 
[Fig.~4(a), left side]. For an ion with the lower tetragonal symmetry of the 
Cu sites in the cuprates, both the  $e_g$ and $t_{2g}$ states are split, as 
illustrated on the right side of Fig.\,4(a).  For a sufficiently large 
tetragonal elongation, the $d_{3z^2-r^2}$ orbital can actually cross the $t_{2g}$ 
manifold and become the lowest-energy orbital, a situation encountered in SCOC 
and in the extreme case of the infinite-layer material CaCuO$_2$, where the 
apical O anions are missing \cite{Moretti}. Thus three peaks are expected, 
and usually observed, in cuprate materials. By contrast, the $dd$ manifold 
of Ba-2342 contains four peaks, visible in the representative spectrum shown 
in Fig.\,4(b). The extra peak reflects the presence of the two inequivalent 
Cu sites (Cu$_A$ and Cu$_B$) in the crystal structure. The two sites have the 
same in-plane coordination, but Cu$_A$ ions have two apical Cl ions with a 
separation $d_{Cu-Cl} = 3.43$\,\AA, while Cu$_B$ ions have two apical Ba ions 
with a much larger separation $d_{Cu-Ba} = 4.94$\,\AA.

Table\,I summarizes, for several representative cuprates, the in-plane Cu-O 
and out-of-plane Cu-anion separations, as well as the energetic separations 
of the different crystal-field levels, as extracted from RIXS data. Quite 
generally. for a given in-plane separation, changes in the apical separation 
have a large effect on the energy of the $d_{3z^2-r^2}$ state but a relatively 
small effect on the energies of the $t_{2g}$ manifold. For Ba-2342 we may 
therefore expect equal energies for the $d_{xy}$ and $d_{xz,yz}$ peaks at both 
the Cu$_A$ and Cu$_B$ sites, but well-separated $d_{3z^2-r^2}$ peaks. The line 
shape illustrated in Fig.~4(b) can indeed be reproduced rather well by four 
peaks. From this procedure, which we assume to include both point-charge 
effects and covalency contributions, we obtain an accurate fit of the 
effective tetragonal crystal-field parameters required to obtain the 
measured $3d$ level splittings \cite{Bersuker}. For Cu$_A$ we obtain the 
parameters $10\,Dq = 1.57$ eV, $D_s = 0.32$ eV, and $D_t = 0.15$ eV while for 
Cu$_{B}$ $10\,Dq = 1.57$ eV, $D_s = 0.37$ eV, and $D_t = 0.18$ eV. The measured 
$dd$ spectra have an angular dependence of their intensities, which is shown 
in Fig.\,5, and also a polarization dependence. These features can be 
reproduced by varying the relative intensities of the four peaks, in a manner 
consistent with the angular dependence of the cross-section predicted by a 
single-ion model \cite{Moretti}, but we do not pursue these details here. 

Two aspects of the analysis are worth noting: i) a good fit is obtained with 
Gaussian line shapes and ii) the energy width of the peaks increases with 
the energy loss. After deconvolving the experimental resolution we obtain 
peak widths (FWHM) of $170$\,meV for $3d_{xy}$ and $3d_{xz,yz}$ and $270$\,meV 
and $380$\,meV, respectively, for the Cu$_A$ and Cu$_B$ $3d_{3z^2-r^2}$ orbitals.
The increase of line width with energy loss is consistent with a progressively 
shorter life time of the excited hole state, but pure life-time broadening 
should yield a Lorentzian contribution. The Gaussian line shape suggests an 
underlying dispersion of these excitonic features, which could be either of 
purely electronic origin, or possibly assisted by phonons. The spin-flip and 
non-spin-flip final states are degenerate for the $3d_{xy}$ and $3d_{xz,yz}$ 
orbitals, but not for the $3d_{3z^2-r^2}$ states, and therefore spin splitting 
can also contribute to the line width of the latter \cite{Moretti}.

\begin{figure}[t]
\includegraphics[scale=0.8]{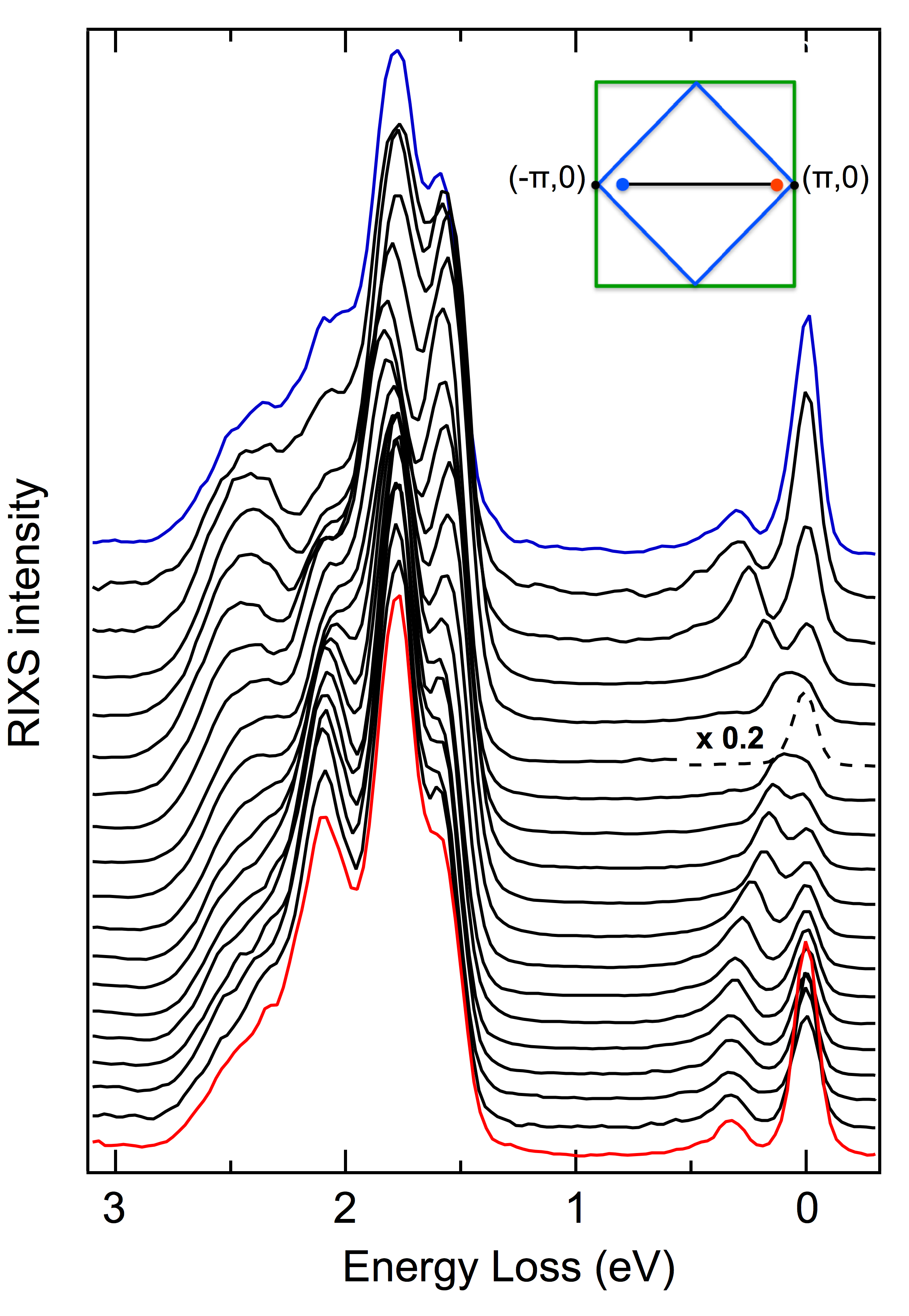}
\caption{Cu L$_3$ RIXS spectra covering nearly the whole Brillouin zone along 
the high-symmetry direction from $(-\pi,0)$ to ($\pi$,0), as shown in the 
inset. The momentum interval between pairs of spectra is not uniform, with 
spectra being denser for positive $q$ values. At the specular-reflection 
condition, ${\bf q} = (0,0)$, the large elastic peak has been reduced by a 
factor of 5 for clarity.}
\label{fig5}
\end{figure}

\subsection{Magnetic excitations}

Figure 5 presents RIXS data for Ba-2342 spanning most of the BZ along the 
high-symmetry direction from $(-\pi,0)$ to $(\pi,0)$, which corresponds to 
the direction of the Cu-O-Cu bond. The data at all angles exhibit the 
characteristic RIXS spectra of cuprates \cite{Guarise}, as discussed in 
Figs.~3(c) and 4. Here we focus on the magnetic spectral features, which 
are also generic in cuprates, namely a resolution-limited peak dispersing 
symmetrically from $(0,0)$ and a broad tail at higher energies. Both features 
are shown in full detail in Fig.\,6(a). The main peak, which is the single 
magnon, disperses up to an energy of $0.32$\,eV near the $(\pm\pi,0)$ zone 
boundaries. The high-energy tail, which extends to approximately $0.6$\,eV, 
corresponds to the multi-magnon continuum. The momentum dependence along 
this cut is also typical of the insulating cuprate parent compounds, such 
as SCOC \cite{SaraReview}.

\begin{figure*}
\begin{center}
\includegraphics[width=\textwidth]{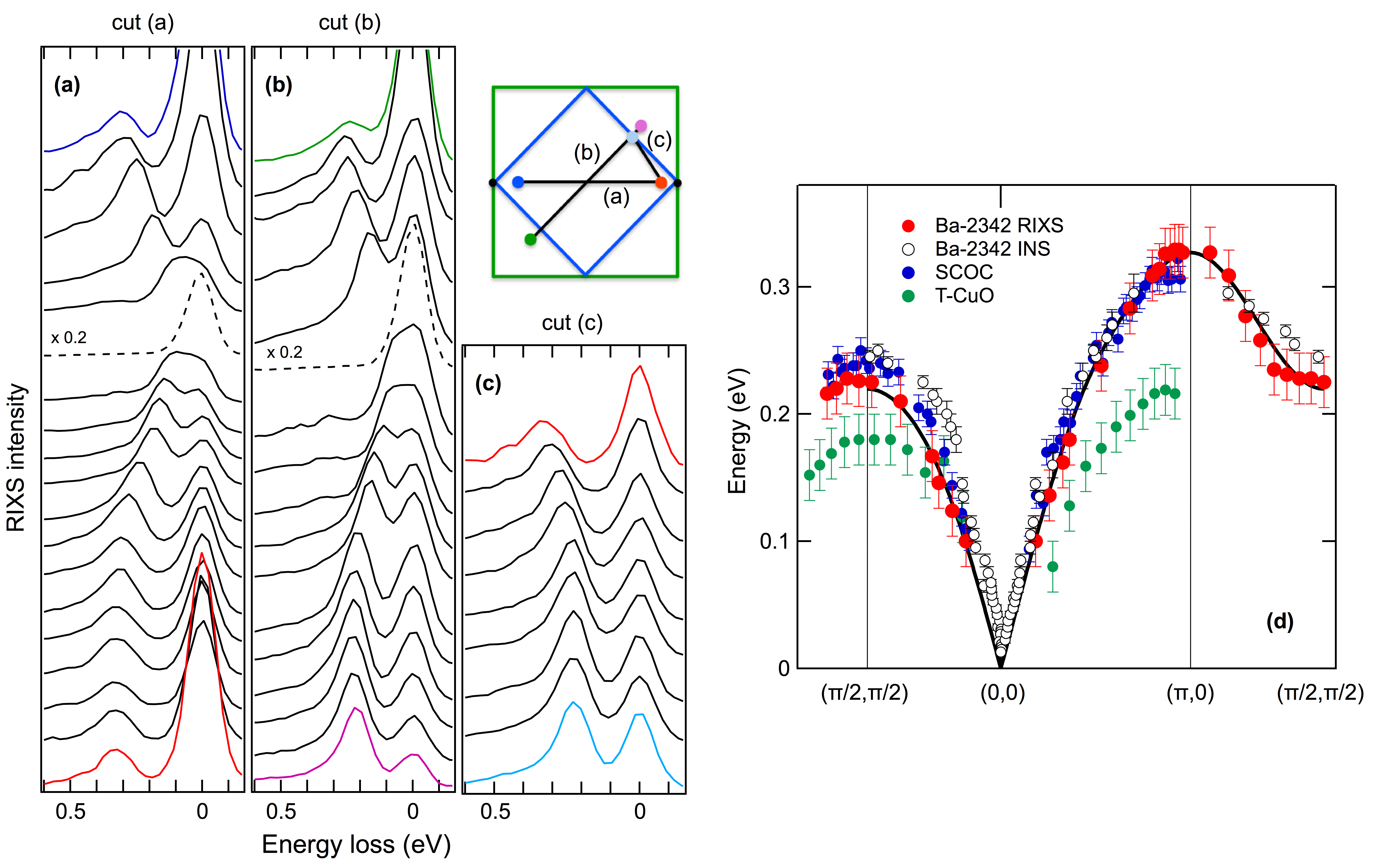}
\caption{Dispersion of the magnetic excitations along (a) the direction 
from $(-\pi,0)$ to $(\pi,0)$, denoted as cut (a) in the inset (these data 
show in detail the low-energy regime of Fig.\,5); (b) the direction from 
$(-\pi,-\pi)$ to $(\pi,\pi)$ [cut (b)], and (c) the direction from $(\pi/2,
\pi/2)$ to near $(\pi,0)$ [cut (c)]. The green and blue squares in the inset 
are respectively the structural and AFM Brillouin zones of the Cu$_A$ 
sublattice, as in Fig.\,1(d). (d) Summary of magnon dispersion data in Ba-2342 
(red symbols) as extracted from panels (a)-(c). For comparison are shown the 
RIXS dispersion relations obtained for SCOC in Ref.~\cite{Guarise} 
(blue) and for T-CuO in Ref.~\cite{Moser2015} (blue). Open symbols mark 
the magnon dispersion obtained by INS in Ref.\,\cite{Babkevich}. The 
solid black line is the dispersion calculated with the model described in 
the text, for the parameters shown in Tables II and III.}
\label{fig6}
\end{center}
\end{figure*}

Further quantitative analysis of the magnon peak reveals that its intensity 
is reduced on approaching the BZ boundaries. Near $(\pi,0)$, in the bottom 
part of Fig.~6(a), the spectra were measured at near-grazing emission, and the 
intensity attenuation is due to self-absorption. At the opposite end, however, 
the spectra were measured near normal emission, where self-absorption is 
negligible. A similar anomalous intensity reduction was already observed 
by Braicovich {\em et al.}~in L$_3$ RIXS data on La$_2$CuO$_4$ (LCO) 
\cite{Braicovich} and was attributed tentatively to quantum corrections to 
the spin waves. Such corrections, predicted by theory, have been verified 
quantitatively by INS in the low-$J$ material copper deuteroformate 
tetradeurate (CFTD), which constitutes a model $S = 1/2$ square-lattice 
antiferromagnet \cite{Christensen,DallaPiazzaNature,Ronnow}.

Figure 6(b) presents data for the BZ diagonal. The magnon dispersion 
reaches a maximum of $E = 0.23$\,eV at the $(\pm\pi/2,\pm\pi/2)$ boundaries 
of the AFM BZ of the Cu$_A$ sublattice. Figure 6(c) illustrates the dispersion 
between $(\pi/2,\pi/2)$ and $(0.95\pi,0)$, along a line (inset) that almost 
coincides with the magnetic zone boundary. The zone-boundary dispersion is of 
particular significance in the theoretical interpretation of the magnetic 
excitation spectrum. In the simplest version of the 2D Heisenberg model, which 
considers only NN interactions, and in a linear spin-wave theory, the magnon 
energy should be constant along the zone boundary. A dispersion along this cut 
therefore indicates longer-range magnetic interactions or magnon interactions, 
or both. Coldea {\em et al.}~observed a dispersion of some $25$\,meV along the 
magnetic BZ of LCO by INS measurements \cite{Coldea}, and fitted their data 
with an extended Heisenberg model that included a four-spin ring-exchange 
interaction. A larger zone-boundary dispersion, of 70\,meV, was observed by 
RIXS measurements in SCOC \cite{Guarise}, and this was reproduced using an 
extended $t$-$t'$-$t''$-$U$ Hubbard model, from which an effective spin 
Hamiltonian and a hierarchy of magnetic couplings can be derived 
\cite{DallaPiazza}. From Fig.~6(c) we find that the magnon dispersion along 
the magnetic zone boundary in Ba-2342 is quite similar to that of SCOC, and 
we pursue its theoretical analysis in Sec.~IIIC. 

The data of Figs.~6(a)-6(c) are summarized in Fig.\,6(d), which shows in 
red the magnon dispersion data along the high-symmetry directions of the BZ. 
The first important conclusion to be drawn is that the A-sublattice dispersion 
follows the symmetry of the magnetic BZ of a single CuO$_2$ sublattice (the 
blue square in the inset of Fig.~6). In particular, for Ba-2342 we do not 
observe a backfolding at $(\pi/2,0)$, which is the boundary of the magnetic 
BZ for the entire Cu$_3$O$_4$ plane [red square in Fig.\,1(d)], even though 
the data were collected at $20$\,K, where both the Cu$_A$ and Cu$_B$ 
sublattices are magnetically ordered. This is a strong indication that 
the two sublattices are not strongly coupled. 

A valuable guide to understanding the RIXS dispersion of Ba-2342 [Fig.~6(d)] 
is provided by comparing with the RIXS dispersion of SCOC \cite{Guarise} 
(blue circles) and of T-CuO \cite{Moser2015} (green circles), as well as 
with the INS dispersion measured for Ba-2342 in Ref.\,\cite{Babkevich} 
(open circles). Clearly the magnon energies in Ba-2342 are much closer to 
those of SCOC, over the entire BZ, than they are to the dispersion measured 
in T-CuO. This observation provides important information concerning the 
nature of the magnetic interactions in the Cu-O planes of all three materials, 
as we discuss in detail in Sec.~IIIC.

Concerning a comparison between the RIXS and INS results, in fact 
the two methods are strongly complementary in cuprates. INS in Ba-2342 
is not handicapped significantly by sample sizes (the authors of 
Ref.\,\cite{Babkevich} used 8 g of coaligned crystals). However, the 
dispersion along the magnetic zone boundary is generically difficult to 
extract because the scattering intensities here are small. By contrast, the 
200--300 meV energy scales of these magnons in cuprates, as well as their 
wave vectors, have a robust RIXS cross-section. A further challenge to INS 
is the presence of some dispersionless excitations around 300 meV arising 
from water in the glue used to fix the multi-crystal samples, and in fact 
there are no reliable INS data around the ($\pi$,0) point of the BZ. Further, 
RIXS is more sensitive than INS to a hybridization between the high- and 
low-energy (A and B) sublattices that would create the folded high-energy band, 
although as noted above this still could not be detected in our measurements. 
On the other hand, the resolution of modern RIXS instruments remains too large 
to probe the low-energy mode associated with the Cu$_B$ sublattice, and any 
understanding of this feature requires INS measurements (which established a 
$20$\,meV dispersion, also with a periodicity corresponding to the smaller 
magnetic BZ \cite{Babkevich}). Finally, RIXS measurements around the ($\pi$,0) 
point display a significant continuum tail at higher energies, with a 
corresponding reduction of peak amplitude [Fig.~6(c)]; in the absence of a 
quantitative theoretical model for this effect, a conventional fit can cause 
RIXS to extract a higher estimate of the magnon energy than INS [Fig.~6(d)], 
as documented most clearly in Ref.~\cite{Plumb}. With these points in 
mind, we turn now to a theoretical analysis of the magnon dispersion in 
Ba-2342. 

\subsection{Model}

As stated in Sec.~I, coherent efforts are under way in the field of cuprate 
research to obtain estimates of the electronic parameters in each material 
that are consistent across many experimental techniques. This undertaking 
proceeds of necessity in parallel with advances in theoretical methods 
allowing quantitative calculations of the relevant parameters from models 
with decreasing levels of approximation (specifically, magnetic-only 
models, those based on one- or three-band Hubbard models, numerical 
schemes for an exact accounting of correlation effects in clusters).   
Because of the intermediate (half-stuffed) nature of Ba-2342, which provides 
two different types of CuO$_2$ lattice, our results contribute experimental 
information of particular value for this process.

To analyze our results for the magnon dispersion relation, we follow
an approach developed to model RIXS measurements of the magnetic excitations
in other insulating cuprates \cite{DallaPiazza}. This description is based on
a reduction of the microscopic model for the cuprate plane to an effective
one-band Hubbard model, which yields significant electron-transfer (hopping)
terms between first- ($t$), second- ($t'$), and third-NN ($t''$) Cu ions 
(Fig.\,7), and was found to give an excellent account of the spin excitations 
in a number of conventional (unstuffed) cuprate parent compounds 
\cite{DallaPiazza}. 

To model the half-stuffed Ba-2342 system, we introduce in addition the 
transfer term $t'_3$, as defined in Fig.\,7, which corresponds to NN hopping 
on the larger Cu$_B$ sublattice, and comment that two different second-NN 
terms are now required on the Cu$_A$ sublattice, $t'_1$ and $t'_2$, depending 
on the presence of a B-sublattice ion. However, to simplify the discussion of 
magnetic exchange, we set $t_1' = t_2' = t'$, a near-equality being expected 
on the grounds that the dominant microscopic cross-plaquette paths avoid the 
central Cu$_B$ site. Finally, a complete description also requires a hopping 
term, $t_d$, connecting the two sublattices, but for reasons explained in 
detail below we proceed directly to an exchange interaction, $J_d$, for this 
function. 

\begin{figure}[t]
\hspace*{-1cm}
\centering
\includegraphics[scale=0.25]{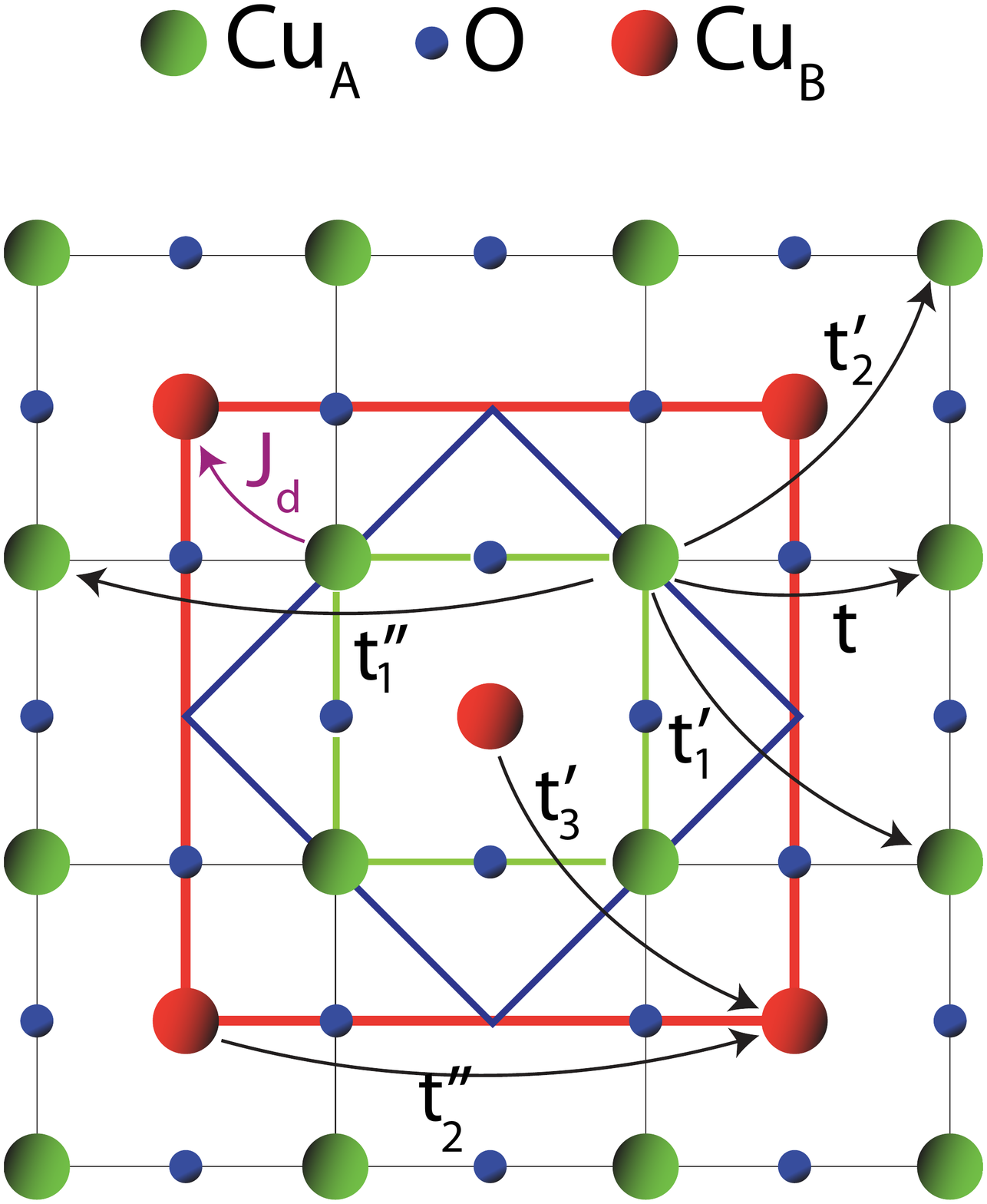}
\caption{Definition of parameters used to model the cuprate plane in Ba-2342. }
\label{fig7}
\end{figure}

We derive \cite{Delannoy} a spin-only Hamiltonian with two-, three- and 
four-spin interactions, retaining terms up to order $t^4/U^3$, where $U$ 
is the on-site Coulomb interaction. To fit the single-magnon dispersion 
obtained in our RIXS measurements, we reduce this further to the effective 
two-spin Hamiltonian
\begin{align}
\mathcal{H} = & \sum_{i,j \in {\rm A}}J^{\rm A}_{n} \mathbf{S}^{\rm A}_i \cdot
\mathbf{S}^{\rm A}_j + \sum_{i,j \in {\rm B}}J^{\rm B}_{n} \mathbf{S}^{\rm B}_i \cdot
\mathbf{S}^{\rm B}_j \nonumber \\
& + \sum_{i,j \in {\rm A,B}}J_{\rm d} \mathbf{S}^{\rm A}_i \cdot \mathbf{S}^{\rm B}_j,
\label{eq:Ham}
\end{align}
where $i \in A$ denotes summation over the $i$th Cu$_A$ site. Here the three- 
and four-spin interactions resulting from the Hubbard model are decoupled into 
additional effective two-spin terms, which renormalize the near-neighbor 
parameters and generate additional terms of longer range. All coefficients 
$J_n^A$ and $J_n^B$ are derived consistently from the Hubbard model of Fig.~7. 
However, the $J_d$ term cannot be derived from a one-band treatment and is 
taken as a free parameter. The spin Hamiltonian in Eq.~(\ref{eq:Ham}) was 
diagonalized using the SpinW library \cite{Toth}. The single-magnon energy 
was renormalized uniformly by a momentum-independent factor, $Z_c = 1.18$ 
\cite{Singh,Igarashi}, and the optimal fit to the dispersion is shown as 
the black solid line in Fig.~\ref{fig6}(d). 

Clearly the theoretical framework provides an excellent fit of the measured 
(A-sublattice) magnon dispersion. Because our description proceeds from the 
Hubbard model, the fitting parameters are $U$, $t$, $t'$, and $t''$; the 
additional parameters $t'_3$ and $J_d$ (Fig.~7) require a knowledge 
of the B-sublattice dispersion and are discussed below. In practice we obtain 
a more statistically reliable fit by fixing $U$, which is known to be largely 
insensitive to precise structural details, to the value $U = 3.5$ eV common 
to all cuprates. The A-sublattice hopping parameters obtained from the fit 
are specified in Table \ref{tab:tU}, where they are compared with the results 
for SCOC and T-CuO. The effective spin interactions, $J^A_n$ and $J^B_n$, are 
obtained as an intermediate step in the process and their values, shown in 
Table~\ref{tab:effJ}, provide helpful physical insight. We comment that these 
parameters are deduced from the full Hubbard model, not from a direct fit, 
and so there is no contradiction in their size, number, or the fact that they 
are subject only to relative errors. Also clear from Fig.~6(d) is that the 
A-sublattice dispersion in Ba-2342 is qualitatively identical to that of 
other cuprate materials and quantitatively similar to that of SCOC, which
is borne out by the close similarity of the Hubbard-model and effective 
spin interaction parameters for the two systems (Tables \ref{tab:tU} and 
\ref{tab:effJ}). This degree of quantitative accuracy serves as a benchmark 
for more detailed theoretical modelling, as we discuss in greater detail below. 

\begin{table}[b]
\begin{tabular}{l|c|c|c|c}
   & $U$\,(eV) & $t$\,(eV) & $t'$\,(eV) & $t''$\,(eV) \\
  \hline
  \hline
 SCOC & 3.5 & 0.480 & $-0.2$ & 0.075 \\
 T-CuO (1) & 3.5 & 0.425 & $-0.2$ & 0.075 \\
 T-CuO (2) & 3.5 & 0.490 & $-0.2$ & 0.075 \\
 Ba-2342 & 3.5 & $\,\, 0.475(11) \,\,$ & $\,\, -0.181(9) \,\,$ & $\,\, 
0.087(4) \,\,$ \\
 \hline
\end{tabular}
\caption{Parameters of the effective one-band Hubbard-model description,
represented in Fig.~7, used to fit the magnon dispersions measured by 
RIXS in SCOC, T-CuO, and Ba-2342. The two parameter sets shown for T-CuO 
correspond to the two scenarios discussed in the text; the complete model 
of scenario (2) also contains an inter-sublattice hopping term, 
$t_d = 0.167$ eV \cite{Moser2015}. The additional parameters in Fig.~7 
that are required to model Ba-2342, $t'_3 = - 0.086$ eV and $J_d = 10$ meV, 
are fixed by the low-energy B-sublattice band and are obtained from INS 
\cite{Babkevich}. 
\label{tab:tU}}
\end{table}

\begin{figure}[t]
\includegraphics[scale=0.6]{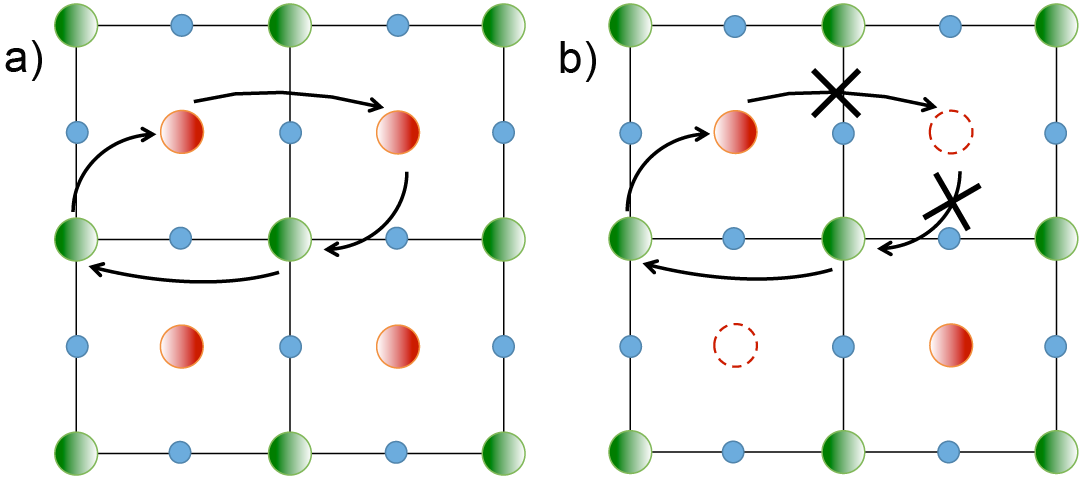}
\caption{(a) Schematic representation of the four-spin exchange process in
T-CuO, which contributes an effective FM interaction between two NN Cu ions
in each sublattice. (b) In the half-stuffed Cu$_3$O$_4$ plane of Ba-2342,
these terms are absent and the A-sublattice interactions are close to those
of the unstuffed (SCOC) system.}
\label{fig8}
\end{figure}

\begin{table}[b]
\begin{tabular}{l|ccccc}
$n$ & 1 & 2 & 3 & 4 & 5\\[0.5em]
\hline
\hline
$J^A_n$ (meV) & 165(5) & 20(4) & 32(2) & 6.2(6) & $-0.20(3)$ \\[0.5em]
$J^B_n$ (meV) & 8.4   & 0.0  & -    & -   & -\\
\hline
\end{tabular}
\caption{Effective spin exchange parameters for the Cu$_A$ and Cu$_B$
sublattices in Ba-2342, obtained from the one-band Hubbard-model parameters 
of Table \ref{tab:tU} and used in Eq.~(\ref{eq:Ham}). We state for clarity 
that the interaction $J_4$ connects a site to its A-sublattice neighbors at 
positions ($\pm 2,\pm 1$) and ($\pm 1,\pm 2$) while $J_5$ connects a site to 
those at ($\pm 2,\pm 2$).
\label{tab:effJ}}
\end{table}

Turning to the question of achieving a convergence of electronic parameters 
between INS and RIXS, the same modelling procedure was used to describe the 
INS measurements on Ba-2342 and the results of Tables~\ref{tab:tU} and 
\ref{tab:effJ} are to be compared with Tables II and III of 
Ref.~\cite{Babkevich}. As discussed above, RIXS and INS are 
complementary methods when applied to cuprates. At high energies, where 
appreciable differences are visible in the magnon dispersion along the zone 
boundary [Fig.~6(d)], RIXS is statistically more reliable but remains subject 
to its coarser energy resolution. A larger band width is observed by RIXS, due 
not only to the higher ($\pi$,0) energy (see above) but also a lower ($\pi$/2,
$\pi$/2) energy, and this is reflected in larger $J_1^A$ and particularly 
$J_3^A$ terms. At the level of the Hubbard model, only the RIXS $t'$ term is 
smaller, while $t$ and $t''$ are identical within the error bars, i.e.~the 
overall parameter sets are manifestly very similar. At low energies, INS is 
more reliable, and indeed the present analysis is compelled to neglect 
out-of-plane coupling and anisotropic exchange interactions, because these 
modify the magnetic spectrum on an energy scale below 10 meV \cite{Babkevich}, 
which is well below the resolution of RIXS. The effective model of 
Eq.~(\ref{eq:Ham}) does include the magnetic fluctuations on the Cu$_B$ 
sublattice, although again the value $t'_3 = - 0.086$ eV required to account 
for these excitations, which are not resolvable in the present experiment, 
must be obtained from INS. Finally, the near-coincidence of the SCOC and 
Ba-2342 A-sublattice magnon bands indicates empirically that the 
inter-sublattice magnetic coupling is very weak, and indeed it is best 
fitted by a ferromagnetic interaction $J_{d} = - 10$ meV \cite{Babkevich}. 
The dominant effect of this term is to produce a small spin gap at the 
magnetic zone center, $(0,0)$, whose size constrains this coupling very 
accurately, giving a rather small value that is only 6\% of $J^A_1$.

Comparison between the RIXS data of Fig.\,6(d) and the result of our
model calculations also allows us to address the evolution of the effective
magnetic interactions with the stoichiometry of the Cu-O planes. We stress
again that the corner-sharing CuO$_2$ sublattice provides the common framework
in which to describe the measured dispersion relations. We have previously
proposed two alternative scenarios, summarized in Table \ref{tab:tU}, to 
explain the significant band-width reduction, of order 30\%, between SCOC and 
T-CuO \cite{Moser2015}, where it is clear from the splitting of the symmetric 
and antisymmetric magnon bands that $J_d$ is also very small. One scenario was 
a reduction of the Cu-O hybridization, which would appear a reduction of the 
first-NN hopping parameter, $t$, in the effective Hubbard Hamiltonian [Table 
\ref{tab:tU}, T-CuO (1)]. The other was that the near-neighbor hopping terms 
remain essentially unchanged [Table \ref{tab:tU}, T-CuO (2)] and the reduction 
arises as the consequence of an effective four-spin inter-sublattice 
interaction of the type represented in Fig.~\ref{fig8}(a), which in the 
Hubbard-model analysis leads to a FM contribution $- 160\,t^2 t_d^2 / U^3$ 
that appears in the model of Eq.~(\ref{eq:Ham}) as a FM intra-sublattice 
interaction. Clearly one may discard a third possibility, of a band-reduction 
due to the FM $J_d$ term, as this is much too small to have such an effect.

Our results for Ba-2342 provide unequivocal evidence in favor of the second
scenario. At the qualitative level, if stuffing the CuO$_2$ lattice were to
cause a gradual reduction in Cu-O hybridization, one would expect the
single-magnon bandwidth in half-stuffed Ba-2342 to lie approximately halfway
between SCOC and T-CuO, which is not the case. By contrast, the effective 
four-spin interaction is no longer possible in the half-stuffed lattice, as 
depicted in Fig.~\ref{fig8}(b), which would explain why Ba-2342 has essentially 
the same band width as SCOC. At a quantitative level, the interpretation in 
terms of a four-spin process requires that the inter-sublattice hopping term 
should take the value $t_d = 167$ meV extracted for T-CuO \cite{Moser2015}. 
Here it is necessary to state that there is no inconsistency with the value 
of $J_d$ required to fit the magnon bands of Ba-2342. The derivation of $J_d$ 
lies beyond the single-band Hubbard model, mandating the consideration of 
charge-transfer terms and the energy splitting of singlet and triplet states 
of two electrons in orthogonal O$_{2p}$ orbitals \cite{rpwa}. A complete 
verification of this interpretation would require the analysis of an extended 
multi-band model containing the overlap integrals $t_{pd}$ and $t_{pp}$ linking 
the Cu$_A$ $3d_{x^2-y^2}$, Cu$_B$ $3d_{x^2-y^2}$, O $2p_x$, and O $2p_y$ orbitals, 
as well as effective Hubbard terms for Cu and O (in both spin states) and 
charge-transfer terms. Alternatively, {\it ab initio} multi-reference 
configuration interaction calculations could be attempted on clusters of 
several unit cells [Figs.~\ref{fig8}(a) and \ref{fig8}(b)] embedded in a 
surrounding matrix \cite{BabkevichPRL}. We remark in closing that the 
unstuffed, half-stuffed, and fully-stuffed cuprates provide an excellent 
case study for testing the quantitative accuracy of all such first-principles 
techniques, given the precision and accuracy of the structural and dynamical 
information available for these systems.

\section{Conclusion}

In summary, by using Cu L$_3$ edge RIXS we have measured the spin-wave 
dispersion in the ``half-stuffed'' Cu-O planes of Ba-2342. Our results, 
which exploit a quite different probe and different set of cross-sections, 
are complementary to data measured by INS. The experimental measurements are 
reproduced at a semi-quantitative level by considering two independent cuprate 
sublattices. Our data clarify previous RIXS results on tetragonal CuO (T-CuO) 
and demonstrate that the substantial reduction of the magnon bandwidth in 
T-CuO with respect to typical AFM layered cuprates is due to effective 
inter-sublattice four-spin interactions, which operate only in a fully-stuffed 
system and are therefore absent in Ba-2342. Taken together, the experimental 
determination of spin-wave dispersions in the series SCOC, Ba-2342, and T-CuO 
provide an opportunity, complementary to and to a significant extent more 
stringent than the more common ARPES results, for an accurate comparison 
with state-of-the-art theoretical (multi-band or {\it ab initio}) calculations 
of hopping and exchange interactions in the cuprates.

\section{Acknowledgments}

We are grateful to S. Moser for valuable discussions. This work 
was supported by the Swiss NSF. The SAXES instrument at the ADRESS 
beam line of the Swiss Light Source was built jointly by the Paul Scherrer 
Institut, EPFL, and Politecnico di Milano. J.P. and T.S. acknowledge 
financial support through Dysenos AG by Kabelwerke Brugg AG Holding and the 
Fachhochschule Nordwestschweiz. J.P. acknowledges financial support by the
Swiss NSF Early Postdoc Mobility fellowship Project No.~P2FRP2-171824. X.L. 
acknowledges financial support from the EU Seventh Framework Programme 
(FP7/2007Ð2013) under Grant No.~290605 (COFUND: PSI-FELLOW) and the Swiss 
NSF through its Sinergia network ``Mott Physics Beyond the Heisenberg Model'' 
(MPBH).


\bibliographystyle{apsrev4-1}

\end{document}